\documentclass[modern]{aastex631}
\usepackage{showyourwork}
\usepackage{amsmath,esint}
\usepackage{acro}

\usepackage{listings}
\usepackage{xcolor, fontawesome}
\usepackage{multirow}
\usepackage{xspace}
\usepackage{hyperref}
\definecolor{mygreen}{rgb}{0,0.6,0}
\definecolor{mygray}{rgb}{0.5,0.5,0.5}
\definecolor{mymauve}{rgb}{0.58,0,0.82}
\definecolor{linkcolor}{rgb}{0.1216,0.4667,0.7059}

\acsetup{
  single    = false,
  list/sort  = true,
  cite/group = true,
  cite/group/cmd = \citealt,
  cite/group/pre = {\xspace; },
  patch/longtable=false
}

\DeclareAcronym{iso}{
  short = ISO,
  long = International Organization for Standardization,
}

\DeclareAcronym{gpu}{
  short = GPU,
  short-plural=s,
  long = Graphics Processing Unit,
  long-plural = s
}

\DeclareAcronym{lsd}{
  short = LSD,
  long = Least-Squares Deconvolution
}

\DeclareAcronym{ALP}{
  short = ALP,
  long = associated Legendre polynomial,
}

\DeclareAcronym{hsh}{
  short = HSH,
  long = hemispherical harmonics
}

\DeclareAcronym{chsh}{
  short = CHSH,
  long = complementary hemispherical harmonics,
  cite = {zheng2019}
}

\DeclareAcronym{osa}{
  short = OSA,
  long = Optical Society of America,
}

\DeclareAcronym{ansi}{
  short = ANSI,
  long = American National Standards Institute,
}
\DeclareAcronym{fim}{
  short = FIM,
  long = Fisher Information Matrix,
  cite = {fisher1922}
}

\DeclareAcronym{hmc}{
  short = HMC,
  long = Hamiltonian Monte Carlo,
  cite = {Betancourt2017}
}

\DeclareAcronym{mcmc}{
  short = MCMC,
  long = Markov Chain Monte Carlo,
  cite = {Metropolis1953}
}

\DeclareAcronym{crlb}{
  short = CRLB,
  long = Cram\'{e}r-Rao Lower Bound,
  cite = {rao1945,cramer1946}
}

\DeclareAcronym{mas}{
  short = mas,
  long = milliarcseconds,
}

\DeclareAcronym{snr}{
  short = SNR,
  long = signal-to-noise ratio
}

\DeclareAcronym{jit}{
  short = jit,
  long = Just-In-Time
}

\DeclareAcronym{hpc}{
  short = HPC,
  long = High-Performance Computing
}
\DeclareAcronym{chara}{
  short = CHARA,
  long = Center for High Angular Resolution Astronomy, 
}
\DeclareAcronym{nir}{
  short = NIR,
  long = near-infrared, 
}
\DeclareAcronym{veritas}{
  short = VERITAS,
  long = Very Energetic Radiation Imaging Telescope Array System, 
}
\DeclareAcronym{iact}{
    short = iACT,
    long = imaging atmospheric-Cherenkov telescope,
}

\DeclareAcronym{cta}{
    short = CTA,
    long = Cherenkov Telescope Array,
}

\DeclareAcronym{mse}{
  short = MSE,
  long = mean squared error,
}

\DeclareAcronym{api}{
  short = API,
  long = Application Programming Interface,
}

\DeclareAcronym{dftm}{
  short = DFT,
  long = discrete Fourier Transform,
}

\DeclareAcronym{tess}{
  short = TESS,
  long = Transiting Exoplanet Survey Satellite,
  cite = {ricker2014tess}
}

\DeclareAcronym{ffi}{
  short = FFI,
  long = Full Frame Image,
}

\DeclareAcronym{toliman}{
  short = TOLIMAN,
  long = Telescope for Orbit Locus Interferometric Monitoring of our Astronomical Neighbourhood,
}
\begin{document}

\title{Analytic Interferometry of Rotating Stellar Surfaces}

\author{\href{https://orcid.org/0000-0001-9145-8444}{Shashank Dholakia}} 
\affiliation{School of Mathematics and Physics, University of Queensland, St Lucia, QLD 4072, Australia}

\author{\href{https://orcid.org/0000-0003-2595-9114}{Benjamin J. S. Pope}}
\affiliation{School of Mathematical \& Physical Sciences, Macquarie University, 12 Wally's Walk, Macquarie Park, NSW 2113}
\affiliation{School of Mathematics and Physics, University of Queensland, St Lucia, QLD 4072, Australia}

\begin{abstract}
The surfaces of rotating stars serve as a window into their interiors, magnetic dynamos, and are important in other areas including exoplanet discovery and atmospheric characterization. While indirect techniques such as photometry and Doppler imaging have been studied for their ability to map stellar surfaces, the gold standard remains optical long-baseline interferometry. In this paper, we develop new closed-form solutions for the interferometric visibility of a rotating star with an arbitrary inhomogeneous surface. We introduce the concept of `stellar rotation synthesis' in interferometry--an analog of Earth rotation synthesis--where stellar rotation adds information to the spherical harmonic modes representing the star's surface intensity. We implement these solutions in the open-source package \texttt{harmonix}, written in \textsc{Jax} with automatic differentiation, providing a rich ecosystem for fitting and inference. Inspired by similar studies for photometry and Doppler imaging, we use simulations of a fiducial star as observed by the \acs{chara} Array and intensity interferometers to perform a comprehensive theoretical study of the information theory of the starspot mapping problem in interferometry. We show that adding simultaneous photometry from a space-based instrument such as TESS adds complementary spatial information to interferometry and can improve the precision on the map coefficients by over an order of magnitude, enabling the detailed mapping of nearby main-sequence stars with current facilities. Finally, we evaluate the performance of existing and proposed intensity interferometers for stellar surface mapping.
\href{https://github.com/shashankdholakia/harmonix.git}{\faGithub} \href{https://github.com/shashankdholakia/analytic-interferometry-paper}{\faBarChart}
\href{https://harmonix.readthedocs.io/latest/}{\faBook}
\end{abstract}

\section{Introduction}
\label{sec:intro}
Recent advances in optical long-baseline interferometers have increasingly permitted the imaging and detection of features such as spots and faculae on the surfaces of individual stars. The longest baselines have already produced images of the surfaces of large, evolved stars with active chromospheres \citep{roettenbacher2016, martinez2021}. More recent advances could soon permit the detection of such features even on nearby main-sequence stars \citep{mourard2018, roettenbacher2022}.

Other techniques exist to probe the surfaces of stars outside the Sun that have been used on a wider variety of stars, such as Doppler imaging \citep{vogt1987} and light curve inversion \citep{harmon2000}, including using occultations from companion stars or exoplanets as in \citet{morris2017}. However, these methods can have limitations in their ability to definitively resolve features, demonstrated by \cite{roettenbacher2017}, whose comparative study revealed significant differences in spot maps obtained by the three methods of interferometry, Doppler imaging, and light curve inversion. 

These discrepancies can be understood from the information theory of each method. Doppler imaging and rotational light curve mapping are both understood to be ill-posed problems, where data are consistent with more than one solution. In the case of rotational light curve mapping, there are in fact an infinite number of surface maps that can analytically fit a given light curve \citep[a fact that was first recognized by][]{russell1906}. \citet{khoklova1976} formalized the problem of mapping a star spectroscopically as an integral equation and others \citep{goncharskii1977, piskunov1990} recognized the problem as ill-posed. With both photometric and spectroscopic methods, regularizers such as maximum entropy \citep{narayan1986} or Tikhonov regularization \citep{tikhonov1987} are often used as a means of reducing the space of possible solutions. However, these and other regularizers can also have the unintended consequence of casting one's assumptions onto the solution space, potentially biasing the resultant map by either suppressing or introducing details \citep{piskunov1990b}. 

Further understanding the discrepancies between techniques and the biases introduced through regularization necessitates a theoretical approach to information theory. A formal description of the degeneracies of photometric and spectroscopic stellar surface mapping and their information theory is presented in \citet{luger2021a, luger2021c, luger2021b}. These works provide newer analytic methods to forward model photometric and spectroscopic observations, which permit a description of the information theory behind the mapping problem. 

The analytic, linear frameworks for describing a stellar surface in \citet{luger2021a, luger2021b} rely on describing the stellar surface in terms of the real spherical harmonics. This is in contrast to other methods that discretize the model stellar surface and perform numerical integrations over these pixels to recover observed quantities such as the light curve or \ac{lsd} profiles \citep{vogt1987, harmon2000}. This work also heavily relies on the spherical harmonics to provide such analytic quantities; we therefore provide a brief introduction to them below. 

The spherical harmonics, being the representation for the rotation group on the sphere, are an attractive basis for representing the intensity on a stellar surface, allowing the rotation of a spotted star to be described analytically and efficiently. We define the complex spherical harmonics using the product of an associated Legendre polynomial and an angular factor:

\begin{equation}
Y_{\ell}^{m}(\vartheta,\varphi) =
N_{\ell m} P_{\ell}^{m}(\cos\vartheta)\,e^{i m \varphi}
\end{equation}
\noindent where we use $\vartheta$ as the inclination angle and $\varphi$ as the azimuthal angle using the \acs{iso} convention, and define a normalization factor:

\begin{equation}
N_{\ell m} = \sqrt{\frac{2\ell+1}{4\pi}\,\frac{(\ell-m)!}{(\ell+m)!}}.
\end{equation}

Since we use the spherical harmonics to represent the intensity on the stellar surface, which is always real, we can then define the orthonormal real spherical harmonics \citep[following][]{starry2019}:

\begin{equation}
Y_{\ell m}(\vartheta,\varphi) =
\begin{cases}
\sqrt{2}\,N_{\ell m}\,P_\ell^{m}(\cos\vartheta)\,\cos(m\varphi), & m>0, \\[6pt]
N_{\ell 0}\,P_\ell(\cos\vartheta), & m=0, \\[6pt]
\sqrt{2}\,N_{\ell |m|}\,P_\ell^{|m|}(\cos\vartheta)\,\sin(|m|\varphi), & m<0,
\end{cases}
\end{equation}

Interferometry remains the only direct method for imaging the surfaces of stars which is possible using current technology. Nevertheless, interferometric imaging of stellar surfaces is itself an ill-posed problem. Unlike in standard imaging, a two-telescope interferometer only samples one point in the frequency domain. The incomplete baseline coverage of the interferometer leads to significant gaps in the spatial frequency domain, also known as the $uv$ plane. There is also information added by observing the star at multiple rotational phases \citep{roettenbacher2016}, but incomplete phase coverage or low cadence may also lead to degeneracies. 

In Sec.~\ref{sec:solution}, we show that if a star's surface is described using spherical harmonics, there exists a linear operation with closed-form expressions to compute the visibility function, which is foundational to the observables produced by interferometers. In Sec.~\ref{sec:infotheory}, we introduce the concept of `stellar rotation synthesis' as applied to the interferometry of resolved rotating bodies and describe the information content of such data. In  Sec.~\ref{sec:harmonix}, we describe the implementation of this model in the open-source code package \texttt{harmonix} developed in the \textsc{Jax} framework. Lastly, we discuss the implications and limitations of the methods and paper in Sec.~\ref{sec:discussions} and summarize in Sec.~\ref{sec:conclusions}.

\section{Interferometry in the spherical harmonic basis}
\label{sec:solution}

Suppose that we have a spherical star whose intensity map, projected into a 2D at a specific viewing orientation, is defined as $I(x,y)$. If the 3d surface of a star is represented using the real spherical harmonics, then we can write the specific intensity at a point $(x,y)$ (projected onto the sky, in angular units) on the surface  as:

\begin{equation}
    I(x,y) = \mathbf{\tilde{y}}^\mathsf{T}(x,y) \ \mathbf{R} \ \mathbf{y}
\end{equation}

where $\mathbf{\tilde{y}}(x,y)$ is the spherical harmonic basis  defined as in \citet{starry2019}: 

\begin{equation}
    \label{eq:by}
    = \begin{pmatrix}
        Y_{0, 0} &
        Y_{1, -1} & Y_{1, 0} & Y_{1, 1} &
        Y_{2, -2} & Y_{2, -1} & Y_{2, 0} & Y_{2, 1} & Y_{2, 2} 
        & \dots
    \end{pmatrix}^\mathsf{T}
    \ ,
\end{equation}
$\mathbf{R}$ is the rotation matrix into the correct viewing orientation with the viewer at $+\infty$ along the z axis and $\mathbf{y}$ is the vector of spherical harmonic coefficients. Because spherical harmonics form a complete, orthonormal basis on a unit sphere, any map can be represented using a sufficiently high order expansion in the spherical harmonics. This makes it a natural choice to represent the surfaces of stars, planets and other approximately spherical bodies.

Here we show that it is possible to use the same description of the surface of a star in terms of spherical harmonics to compute analytic observables used in interferometry. Recall that interferometric observations record a quantity called the visibility. The van-Cittert Zernike theorem \citep{vancitttert1934, zernike1938} relates the intensity map of a star to the complex coherence function (also called the complex visibility) using the Fourier transform:

\begin{equation} \label{eq:vcztheorem}
V(u,v) = \oiint\limits_{\mathrm{S}(x,y)} I(x,y) e^{-i(ux + vy)} dS
\end{equation}
where $V$ is the complex visibility at spatial frequency $(u,v)$ and $\mathrm{S}$ is the projected disk of the star as a function of $(x,y)$. Using the spherical harmonic basis, we can write the integral as:

\begin{equation} \label{eq:fourierintegral}
   V(u,v) = \oiint\limits_{\mathrm{S}(x,y)} \mathbf{\tilde{y}}^\mathsf{T}(x,y) \ e^{-i(ux + vy)} \ dS \ \mathbf{R} \ \mathbf{y}
\end{equation} 
where $\mathbf{R} \ \mathbf{y}$ are not dependent on x and y and can therefore be pulled out of the integral. In the case of $u, v = 0$, Eq.~\ref{eq:fourierintegral} reduces to the disk-integrated brightness of a body as it rotates:

\begin{equation} \label{eq:fluxintegral}
   F = \oiint\limits_{\mathrm{S}(x,y)} \mathbf{\tilde{y}}^\mathsf{T}(x,y) \ dS \ \mathbf{R} \ \mathbf{y}
\end{equation}
where $F$ is the observed flux as a star rotates (i.e the light curve). This equation finds a closed-form solution in e.g. \citet{cowan2013} and more generally in the open-source \texttt{starry} package by \citet{starry2019} that, among other things, permits a description of the information content of photometric data.

\newpage
\subsection{Solving the Fourier integral}

To solve the integral in Eq.~\ref{eq:fourierintegral}, we first reparametrize the projected disk of the star in the polar coordinate system with variables $r, \theta$. The spatial frequencies are also reparametrized as $\rho, \phi$:

\begin{equation} \label{eq:polarform}
   V(\rho,\phi) = \int_{0}^{2\pi}\int_{0}^{1} \mathbf{\tilde{y}}^\mathsf{T}(r, \theta) \ e^{-i\rho r\cos{(\phi-\theta)}} \ r dr d\theta \ \mathbf{R} \ \mathbf{y}
\end{equation} 

where the integral is explicitly bounded over a projected unit disk. Here we note that the spherical harmonics can be subdivided into two subspaces, the \ac{hsh} and the \ac{chsh}:

\begin{equation} \label{eq:hsh_chsh_union}
    \mathbf{\tilde{y}} = \mathbf{\tilde{y}}_{\mathrm{HSH}} \cup \ \mathbf{\tilde{y}}_{\mathrm{CHSH}}
\end{equation}

We solve the integral in Eq.~\ref{eq:polarform}, finding two families of solutions. Noting the similarity of the hemispheric harmonics to Zernike polynomials, we find that the solutions to terms belonging to this subset can be written as finite sums of Bessel functions. We find that the solution for each mode belonging to the complementary hemispheric harmonics can be written as a single spherical Bessel function.  We show our solution below for a spotted, rotating star without limb darkening and further elucidate, prove, and extend this result in the following sections.

\begin{equation} \label{eq:fullsolution}
\begin{aligned}
    V(\rho, \phi) &= \begin{pmatrix}
\widehat{\mathbf{\tilde{y}}}^\mathsf{T}_{\mathrm{HSH}} \\
\widehat{\mathbf{\tilde{y}}}^\mathsf{T}_{\mathrm{CHSH}}
\end{pmatrix} \mathbf{R}\mathbf{y} \\
\widehat{\mathbf{\tilde{y}}}_{\mathrm{HSH}}   &= \mathbf{A}\,\frac{\mathbf{J}_{l+1}(\rho)}{\rho} \left\{
    \begin{array}{ll}
    \cos(m \phi) &  m \geq 0 \\
    \sin(|m| \phi) &  m < 0
    \end{array}
    \right. \\
\widehat{\mathbf{\tilde{y}}}_{\mathrm{CHSH}} &= \mathbf{B}\,\frac{\mathbf{j}_{l}(\rho)}{\rho}
\begin{cases}
\cos(m \phi), & m \geq 0 \\
\sin(|m| \phi), & m < 0
\end{cases}
\end{aligned}
\end{equation}

\begin{figure}[ht!]
    \script{spherical_harmonics.py}
    \begin{centering}
        \includegraphics[width=\linewidth]{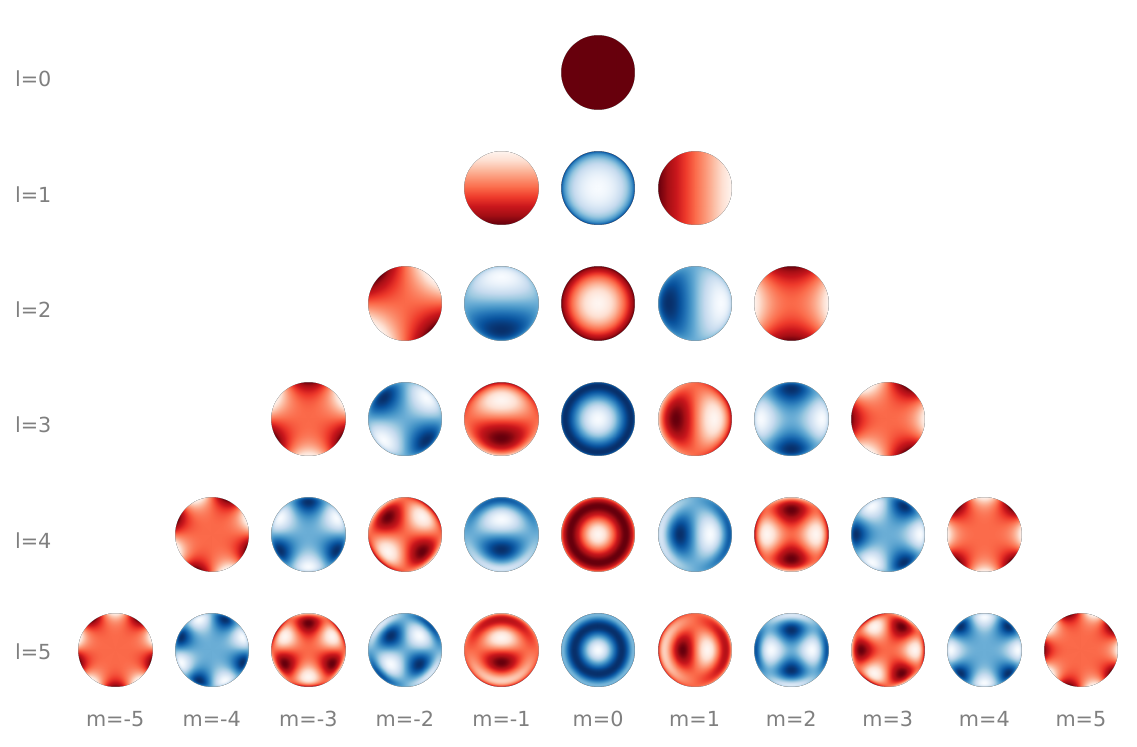}
        \caption{
        Two-dimensional projections of the set of spherical harmonics up to degree $l=5$. The hemispheric harmonics (which are linear combinations of Zernike polynomials) are plotted in red, and the complementary hemispheric harmonics in blue.
\href{https://github.com/shashankdholakia/analytic-interferometry-paper/blob/main/src/scripts/spherical_harmonics.py}{\faGithub}}
        \label{fig:pyramid}
    \end{centering}
\end{figure}

\subsubsection{Hemispheric harmonics} \label{sec:hemisphericharmonics}
The hemispheric harmonics form an orthogonal basis over the visible hemisphere of the star, and also bear a strong resemblance to the Zernike polynomials, which are an orthonormal basis on the unit disk and have a closed-form Fourier transform. 
We can use the \ac{osa} / \ac{ansi} indexing scheme for the Zernike polynomials to define a basis $\mathbf{\tilde{z}}$ where the $j^\mathrm{th}$ element is:

\begin{equation} \label{eq:zernike_j}
    \mathbf{\tilde{z}}_j = R_n^{|m|}(r) \left\{
    \begin{array}{ll}
    \cos(m \theta) &  m \geq 0 \\
    \sin(|m| \theta) &  m < 0
    \end{array}
    \right.
\end{equation}
and where the index $j$ is related to $n, m$ using the relations \citep{niu2022}

\begin{align} j & = \frac{{n(n + 2) + m}}{2}, \nonumber\\ n & = \left\lceil {\frac{{ - 3 + \sqrt {9 + 8j} }}{2}} \right\rceil , \nonumber\\ m & = 2j - n(n + 2),
\end{align}
 Below we show this basis up to $n=3$:
 \begin{equation} \label{eq:zernikebasis}
 \left(\begin{matrix}1\\r \sin{\left(\theta \right)}\\r \cos{\left(\theta \right)}\\r^{2} \sin{\left(2 \theta \right)}\\2 r^{2} - 1\\r^{2} \cos{\left(2 \theta \right)}\\r^{3} \sin{\left(3 \theta \right)}\\\left(3 r^{3} - 2 r\right) \sin{\left(\theta \right)}\\\left(3 r^{3} - 2 r\right) \cos{\left(\theta \right)}\\r^{3} \cos{\left(3 \theta \right)}\end{matrix}\right)
 \end{equation}
\noindent In particular, each $(n, m)$ term in the Zernike basis matches the corresponding $(l, m)$ term of the spherical harmonic basis but with different coefficients to the radial polynomial, where the Zernike radial polynomial $R_n^{|m|}$ is replaced with the associated Legendre polynomial $P_l^{m}(z)$, where we take $z=\sqrt{1-r^2}$ to represent the visible hemisphere of the sphere. Therefore, using the same indexing scheme as for the Zernike basis (see Eq.~\ref{eq:zernike_j}, except replacing $(n, m)$ with $(l, m)$), we can construct a basis we term the hemispheric harmonic basis:

\begin{equation} \label{eq:hsh_basis}
\mathbf{\tilde{y}}_{\mathrm{HSH}, j} = P_l^m(\sqrt{1-r^2}) \left\{
    \begin{array}{ll}
    \cos(m \theta) &  m \geq 0, l+m \text{ even} \\
    \sin(|m| \theta) &  m < 0, l+m \text{ even}
    \end{array}
    \right.
\end{equation}

which we show here up to $l=3$:

\begin{equation} \label{eq:hsh_basis_3}
\left(\begin{matrix}1\\- r \sin{\left(\theta \right)}\\- r \cos{\left(\theta \right)}\\3 r^{2} \sin{\left(2 \theta \right)}\\1 - \frac{3 r^{2}}{2}\\3 r^{2} \cos{\left(2 \theta \right)}\\- 15 r^{3} \sin{\left(3 \theta \right)}\\- \left(- \frac{15 r^{3}}{2} + 6 r\right) \sin{\left(\theta \right)}\\- \left(- \frac{15 r^{3}}{2} + 6 r\right) \cos{\left(\theta \right)}\\- 15 r^{3} \cos{\left(3 \theta \right)}\end{matrix}\right)
\end{equation}
We note that with a finite, square change of basis matrix $\mathbf{A}$, we can transform the Zernike basis into the hemispheric harmonic basis. We start with an intermediate polynomial basis $\tilde{p}$, which consists of a monomial times an angular factor of either $\sin{(|m|\theta)}$ or $\cos{(m \theta)}$ (shown here up to n=3):

\begin{equation}
\left(\begin{matrix}1\\r \sin{\left(\theta \right)}\\r \cos{\left(\theta \right)}\\r^{2} \sin{\left(2 \theta \right)}\\r^{2}\\r^{2} \cos{\left(2 \theta \right)}\\r^{3} \sin{\left(3 \theta \right)}\\r^{3} \sin{\left(\theta \right)}\\r^{3} \cos{\left(\theta \right)}\\r^{3} \cos{\left(3 \theta \right)}\end{matrix}\right)
\end{equation}

If we write the basis change from Zernikes to the polynomial basis as $\mathbf{A}_{\mathrm{Z} \rightarrow \mathrm{P}}$, and from HSH to polynomials as $\mathbf{A}_{\mathrm{HSH} \rightarrow \mathrm{P}}$, then the full basis change is just:

\begin{equation}
    \mathbf{A} = \mathbf{A}_{\mathrm{Z} \leftarrow \mathrm{P}} \ \mathbf{A}_{\mathrm{HSH} \rightarrow \mathrm{P}}
\end{equation}

where $\mathbf{A}_{\mathrm{Z} \leftarrow \mathrm{P}} = \mathbf{A}_{\mathrm{Z} \rightarrow \mathrm{P}}^{-1}$. We note here the Fourier transform of the Zernike basis can be written in terms of Bessel functions of the first kind as: 

\begin{equation} \label{eq:zernike_ft}
 \widehat{\mathbf{\tilde{z}}}_j = (-1)^{n/2-m}\ \frac{J_{j+1}(\rho)}{\rho} \left\{
    \begin{array}{ll}
    \cos(m \phi) &  m \geq 0 \\
    \sin(|m| \phi) &  m < 0
    \end{array}
    \right.
\end{equation}

Because the Zernikes have an analytic Fourier transform, and with the linearity of the Fourier transform, we can write the Fourier transform of the hemispheric harmonics over the visible hemisphere of a body as:
\begin{equation} \label{eq:hsh_to_zern}
  \widehat{\mathbf{\tilde{y}}}_{\mathrm{HSH}, j} = \mathbf{A} \ \widehat{\mathbf{\tilde{z}}}_{j}
\end{equation}

\subsubsection{Complementary hemispheric harmonics}
We find that the other half of the spherical harmonic terms, for which $l+m$ is odd, when simplifying the Legendre polynomial $P_l^m(\sqrt{1-r^2})$, contain an extra factor of $\sqrt{1-r^2}$ and hence do not admit the same transformation into Zernike polynomials as with the hemispheric harmonics. We therefore seek another solution to these terms.

We want to compute the integral:

\begin{equation} \label{eq:polarformchsh}
   V(\rho,\phi) = \int_{0}^{2\pi}\int_{0}^{1} \mathbf{\tilde{y}}^\mathsf{T}(r, \theta) \ e^{-i\rho r\cos{(\phi-\theta)}} \ r dr d\theta \ \mathbf{R} \ \mathbf{y}
\end{equation}

for $l, m$ odd, where we define the complementary hemispheric harmonics as (dropping the normalization factor $A_{lm}$ in the normalized associated Legendre function:

\begin{equation} \label{eq:chsh_basis}
\mathbf{\tilde{y}}_{\mathrm{CHSH}} = P_l^m(\sqrt{1-r^2}) \left\{
    \begin{array}{ll}
    \cos(m \theta) &  m \geq 0, l+m \text{ odd} \\
    \sin(|m| \theta) &  m < 0, l+m \text{ odd}
    \end{array}
    \right.
\end{equation}

\noindent We start by solving the angular part of this integral:

\begin{equation}
    \int_0^{2\pi} e^{-i\rho r \cos(\phi-\theta)} \cos(m\theta)d\theta = 2\pi i^m J_m(\rho r) \cos(m\phi)
\end{equation} 
and 
\begin{equation}
    \int_0^{2\pi} e^{-i\rho r \cos(\phi-\theta)} \sin(|m|\theta)d\theta = 2\pi i^m J_m(\rho r) \sin(|m|\phi)
\end{equation}

\noindent We are then left with (as the radial part of the integral):
\begin{equation}
    I_l^m(\rho) \equiv \int_0^1 P_l^m(\sqrt{1-r^2}) J_m(\rho r) r \,dr
\end{equation}

\noindent Change variables as $z = \cos\theta = \sqrt{1 - r^2}$ or $r = \sin\theta = \sqrt{1 - z^2}$:
\begin{align}
    I_l^m(\rho) &= \int_0^1 P_l^m(z) J_m(\rho \sqrt{1-z^2}) z dz \\
    &= \int_0^{\pi/2} P_l^m(\cos\theta) J_m(\rho \sin\theta) \cos\theta \sin\theta \,d\theta
\end{align}

If $l + m$ is odd, the ALP has odd parity and we can extend the integral to the full sphere:
\begin{align}
    2 I_l^m(\rho) &= \int_{-1}^1 P_l^m(z) J_m(\rho \sqrt{1-z^2}) z dz \\
    &= \int_0^{\pi} P_l^m(\cos\theta) J_m(\rho \sin\theta) \cos\theta \sin\theta \,d\theta
\end{align}

Using the ALP recurrence \citep[12.92]{arfken} to remove the $z$ multiplication:
\begin{align}
    2 I_l^m(\rho) &= \int_{-1}^1 \left( \frac{l-m+1}{2l+1} P_{l+1}^m(z) + \frac{l+m}{2l + 1} P_{l-1}^m(z) \right) J_m(\rho \sqrt{1-z^2}) dz \\
    &= \frac{l-m+1}{2l+1} K_{l+1}^m(\rho) + \frac{l+m}{2l + 1} K_{l-1}^m(\rho)
\end{align}
where we use the relation found in \citep{neves2006}
\begin{align}
    K_l^m(\rho) & \equiv \int_{-1}^1 P_{l}^m(z) J_m(\rho \sqrt{1-z^2}) dz \\
    & = \int_0^\pi P_{l}^m(\cos\theta) J_m(\rho \sin\theta) \sin\theta \,d\theta \\
    &= 2 \frac{(l + m - 1)!!}{(l - m)!!} j_l(\rho) \quad [l+m \; \textrm{even}]
\end{align}

Using the spherical Bessel function recurrence \citep[10.51]{NIST:DLMF}, we then end up with the following:
\begin{align}
    I_l^m(\rho) = \frac{(l + m)!!}{(l - m - 1)!!} \frac{j_{l}(\rho)}{\rho}
\end{align}

\begin{equation} \label{eq:cshs_ft}
\widehat{\mathbf{\tilde{y}}}_{\mathrm{CHSH}, n} = \mathrm{B}_l^m \  \frac{j_l(\rho)}{\rho} \left\{
\begin{array}{ll}
\cos(m \phi) &  m \geq 0 \\
\sin(|m| \phi) &  m < 0
\end{array}
\right.
\end{equation}
\noindent where $B_l^m = 2\pi i^m \frac{(l + m)!!}{(l - m - 1)!!}$
Combining this proof with the result in Section~\ref{sec:hemisphericharmonics}, we have a complete proof for the visibility function of a rotating star as represented in the spherical harmonic basis as presented in Eq.~\ref{eq:fullsolution}.

\subsection{Analytic limb darkening}
Until now, we have described a closed form solution for the complex visibility of arbitrary stellar surfaces when represented in the spherical harmonic basis. Being closed under rotation for a given resolution $l_{\mathrm{max}}$, the spherical harmonics are an ideal basis to represent the surface of rotating stars with inhomogeneities such as spots. Limb darkening, however, is an effect that alters the visible surface of the star without following its rotation. As shown by \citet{starry2019}, this can also be taken into account in our formalism by multiplying the underlying surface map $\mathbf{y}$ with a limb darkening filter after the application of the rotation matrix $\mathbf{R}$.

We represent limb darkening as a radial polynomial similar to many works in exoplanets and stars \citep{quirrenbach1996, agol2020, starry2019} with coefficients $\mathbf{u}$ defined as:
\begin{equation}
   \frac{I(\mu)}{I(1)} = 1 - u_1(1 - \mu) - u_2(1 - \mu)^2 \cdots - u_k(1 - \mu)^k
\end{equation}

Using a change of basis matrix $\mathbf{U}$, we can transform the limb darkening into the spherical harmonic basis, in which only the radial terms ($m=0$) have nonzero coefficients. We now have two sets of spherical harmonic vectors, one representing the rotating component of the stellar surface map, and the other representing the limb darkening. We can then use the spherical harmonic product identity (for instance using Clebsch-Gordan coefficients) to combine these to yield a spherical harmonic representation that is raised in order by the degree of the limb darkening order $k$. This gives us finally:

\begin{equation}
    V(u, v) = \begin{pmatrix}
\widehat{\mathbf{\tilde{y}}}^\mathsf{T}_{\mathrm{HSH}} \\
\widehat{\mathbf{\tilde{y}}}^\mathsf{T}_{\mathrm{CHSH}}
\end{pmatrix} \mathbf{U}\mathbf{u}\mathbf{R}\mathbf{y}
\end{equation}

\section{Information theory}
\label{sec:infotheory}
The use of automatic differentiation provided by \textsc{Jax} in our model (explained in detail in Section~\ref{sec:harmonix}) allows the computation of the Fisher information \citep{fisher1922}, and hence the \ac{crlb}, on relevant parameters of the model, such as the stellar angular diameter, the limb darkening coefficients $\mathbf{u}$ and surface map coefficients $\mathbf{y}$. In a Bayesian interpretation, this is the Laplace approximation -- a variational method approximating the posterior distribution as Gaussian \citep{Kass1991}. In this section, we describe the use of Fisher information in order to make inferences about the information content on the surface maps present in data from existing and planned optical long-baseline interferometers of rotating stars. 

In this section, we use the Fisher information and the related \ac{crlb} for a few different purposes. Firstly, the Fisher information allows us to compare different observational setups to estimate which provides more information on a stellar surface map, and quantitatively compare the performance of existing and planned optical stellar interferometers for the problem of mapping rotating stars. Secondly, the reciprocal of the Fisher information, the \ac{crlb}, provides an estimate of the best-case uncertainty on parameters of interest, such as the stellar angular diameter, limb darkening, or the stellar surface map parameters. This lower bound on uncertainty can be computed from simulated data analytically without resorting to numerical sampling (such as MCMC) to estimate the uncertainty. 

Given an interferometric observation of a rotating, spotted star, we may have a data vector $\mathbf{d}$ and a vector of parameters $\boldsymbol{\boldsymbol{\vartheta}}$ corresponding to latent parameters of the interferometric model; we wish to constrain some of these latent parameters while marginalizing over the other unknown parameters. In our case, the data vector $\mathbf{d}$ could contain interferometric observables such as squared visibilities or closure phases, while the parameter vector could contain certain parameters of interest, such as the surface map $\mathbf{y}$ and other parameters to be marginalized over such as limb darkening coefficients $\mathbf{u}$, stellar inclination and obliquity, etc.

We assume that the observed data are normally distributed about the data vector computed at the true parameters, giving a log-likelihood function:
\begin{equation} \label{eq:loglike}
    \mathcal{L}(\boldsymbol{\vartheta}) \propto -\frac{1}{2} \sum_{i=1}^{N} \frac{(d_i - \mu_i(\boldsymbol{\vartheta}))^2}{\sigma_i^2}
\end{equation}
The log likelihood $\mathcal{L}$, should be maximal at the expected values of these parameters $\boldsymbol{\hat{\vartheta}}$. We can Taylor expand the log-likelihood about the point of maximum likelihood \citep[following][]{Desdoigts2024}:
\begin{equation} \label{eq:taylorfull}
    \mathcal{L}(\vartheta) \approx 
    \underbrace{\mathcal{L}(\hat{\boldsymbol{\vartheta}})}_{\text{0th order}}
    + \underbrace{\nabla \mathcal{L}(\boldsymbol{\vartheta} - \hat{\boldsymbol{\vartheta}})}_{\text{1st order}} + \underbrace{\frac{1}{2}(\boldsymbol{\vartheta} - \hat{\boldsymbol{\vartheta}})^\top \nabla^2 \mathcal{L}(\boldsymbol{\vartheta} - \hat{\boldsymbol{\vartheta}})}_{\text{2nd order}} + \text{higher order terms}
\end{equation}
where in this context, the $\nabla^2$ operator refers to the Hessian matrix, that is:
\begin{equation*}
    \nabla^2 \mathcal{L}(\boldsymbol{\vartheta})_{ij} = \frac{\partial^2 \mathcal{L}}{\partial\vartheta_i\ \partial\vartheta_j}
\end{equation*}

Furthermore, we note that the gradient of $ \mathcal{L}$ will be zero around the point of maximum likelihood: 
\begin{equation}
    \nabla \mathcal{L}(\hat{\boldsymbol{\vartheta}}) = 0
\end{equation}
so we can simplify Eq.~\ref{eq:taylorfull} to read:
\begin{equation} \label{eq:taylor}
    \mathcal{L}(\boldsymbol{\vartheta}) \approx 
    \mathcal{L}(\hat{\boldsymbol{\vartheta}}) + \frac{1}{2}(\boldsymbol{\vartheta} - \hat{\boldsymbol{\vartheta}})^\top \nabla^2 \mathcal{L}(\boldsymbol{\vartheta} - \hat{\boldsymbol{\vartheta}})
\end{equation}

The latter term $\nabla^2 \mathcal{L}$ is also known as the \ac{fim}. Its negative inverse is known as the \acl{crlb}. Both the \ac{fim} and \ac{crlb} have important properties described below which are relevant to the information theory of interferometry:

Firstly, the \ac{fim} is additive, which has the important implication that we can individually construct the \ac{fim} from different sets of observations, for instance, by defining a separate data vector for visibilities $\mathbf{d_{\mathrm{vis}}}$ and closure phases $\mathbf{d_{\mathrm{cp}}}$, and then simply add them to get the total (or joint) information, as might be obtained from a joint fit to both datasets. This can also be done, for example, with data taken on separate nights seeing different faces of the rotating star, or even adding the \ac{fim} from complementary observations such as space-based photometry (see Section.~\ref{sec:simphotometry}. 

One subtlety not mentioned above is that we have assumed that for a given observation with noise, the maximum likelihood yields the true values of the parameters, when in reality it will be at least slightly different. Taking the Hessian of the log-likelihood of a real observation (or simulated, with noise) at the point of maximum likelihood will yield a matrix called the observed information; for an unbiased estimator, the expected value of this observed information for a hypothetical observation is the \ac{fim}. In our work, we are simulating data with a known true parameter vector $\boldsymbol{\vartheta}$ and can compute exactly the \ac{fim} for a hypothetical observation. 

The inverse of the \ac{fim} is a covariance matrix of our parameter vector $\boldsymbol{\vartheta}$, where the diagonals contain a variance on each parameter. Assuming again that the maximum likelihood solution is not biased from the true parameters, this represents a lower bound on the variance, or best-case scenario for the precision obtainable in a set of observations with a given noise, called the \acl{crlb}. 

Aside from its use as a metric for the best-case precision achievable with a certain technique, the \ac{crlb} is also useful in the context of understanding the influence of nuisance parameters on the parameters of interest. After computing the \ac{fim} with all the parameters, the variances of the parameters of interest will naturally be marginalized with respect to nuisance parameters, which can then be dropped from the covariance matrix. 

In the following sections, we simulate data from a star as observed by optical interferometers such as the \ac{chara} Array and \ac{veritas}. We start with the sky position of the fiducial star $\epsilon$ UMa and use \texttt{Skyfield}\footnote{\url{http://rhodesmill.org/skyfield/}} to compute the projected baseline separations over a range of hour angles during a night of observation. We then divide the projected separation by the wavelengths (or in the case of intensity interferometry, the central wavelength of the bandpass), to construct $uv$ tracks on the sky. Using our \texttt{harmonix} model (see Section~\ref{sec:harmonix}), we simulate observables such as visibilities, closure phases, and light curves, from these observations with Gaussian noise representative of the expected precision from the respective instruments. 

\newpage
\subsection{Measuring stellar angular diameters} \label{sec:angdiam}

As a simplified example, let us take the well-known problem in optical stellar interferometry of resolving the angular diameter of a star $\theta$ while marginalizing over the unknown limb darkening $\mathbf{u}$. We first construct the parameter vector for this problem:

\begin{equation}
    \boldsymbol{\vartheta} = 
    \begin{pmatrix} 
    \theta & u_1 & u_2
    \end{pmatrix}^\mathsf{T}
\end{equation}

We then use the procedure described above to simulate a night of observations with the SPICA visible wavelength beam combiner \citep{spica} on \ac{chara} for the same star with 20 different angular diameters. For each of these, we analytically calculate the log-likelihood based on a visibility uncertainty of $1\%$, which allows us to construct the \ac{fim}:

\begin{equation} \label{eq:fim}
\mathbf{FIM} = 
- \begin{pmatrix}
\frac{\partial^2 \mathcal{L}}{\partial \theta^2} & \frac{\partial^2 \mathcal{L}}{\partial \theta \partial u_1} & \frac{\partial^2 \mathcal{L}}{\partial \theta \partial u_2} \\
\frac{\partial^2 \mathcal{L}}{\partial u_1 \partial \theta} & \frac{\partial^2 \mathcal{L}}{\partial u_1^2} & \frac{\partial^2 \mathcal{L}}{\partial u_1 \partial u_2} \\
\frac{\partial^2 \mathcal{L}}{\partial u_2 \partial \theta} & \frac{\partial^2 \mathcal{L}}{\partial u_2 \partial u_1} & \frac{\partial^2 \mathcal{L}}{\partial u_2^2}
\end{pmatrix}
\end{equation}

If we then invert the \ac{fim}, the square root of the top left term of the resulting covariance matrix is the standard deviation on the stellar angular diameter $\sigma_{\theta}$ marginalized over the limb darkening. 

\begin{figure}[ht!]
    \script{chara_sim_ld_radius.py}
    \begin{centering}
        \includegraphics[width=\linewidth]{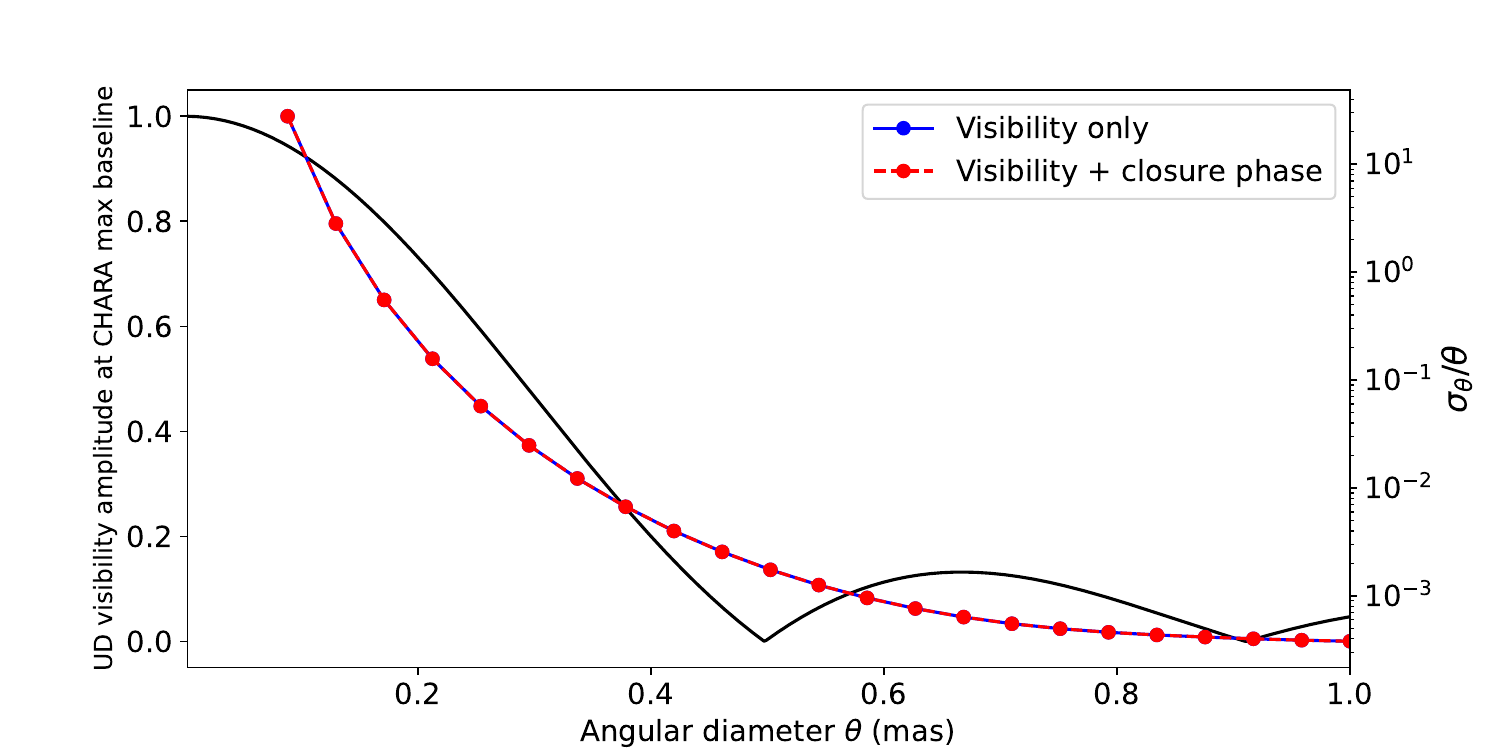}
        \caption{Fractional error on stellar angular diameter for simulated stars ranging from 0.1 to 1.0 milliarcseconds. The black line indicates the visibility curve for a uniform disk star at the baseline of the instrument as a reference; a uniform disk star of a given angular diameter would truncate at that point on the curve. The blue curve shows the fractional error (standard deviation over the angular diameter) with only the visibilities, and the red curve shows the same with visibilities and closure phases. As expected, closure phases provide no additional information in this radially symmetric example. \href{https://github.com/shashankdholakia/analytic-interferometry-paper/blob/main/src/scripts/chara_sim_ld_radius.py}{\faGithub}}
        \label{fig:ldangulardiameter}
    \end{centering}
\end{figure}

\subsection{Stellar rotation synthesis in optical stellar interferometry} \label{sec:rotsynthesis}

In this section, we describe the application of our analytic model to rotating stars with spots, wherein information can be added by observing the star at different rotational phases. These examples are motivated by \citet{vogt1987}, who used the example of a star with the word VOGT to understand the problem of Doppler imaging of rotating stellar surfaces and in particular \citep{luger2021b}, who used an identical setup to elucidate the information theory of mapping stars with Doppler imaging. 

We start with the same $l_{\mathrm{max}}=15$ expansion of an image of the word \texttt{SPOT} as presented in \citet{luger2021b}. The spherical harmonic coefficients represent the intensity on the surface of a star with an inclination of $60^{\circ}$ and an angular diameter of $1.47$ \ac{mas}. As above, we simulate the 6-telescope \ac{chara} Array observing the \texttt{SPOT} star over 8 separate nights, assuming equally-spaced phase coverage and the same 5 hour angles each night. 

\begin{figure}[ht!]
    \script{spot_map_plot_chara.py}
    \begin{centering}
        \includegraphics[width=\linewidth]{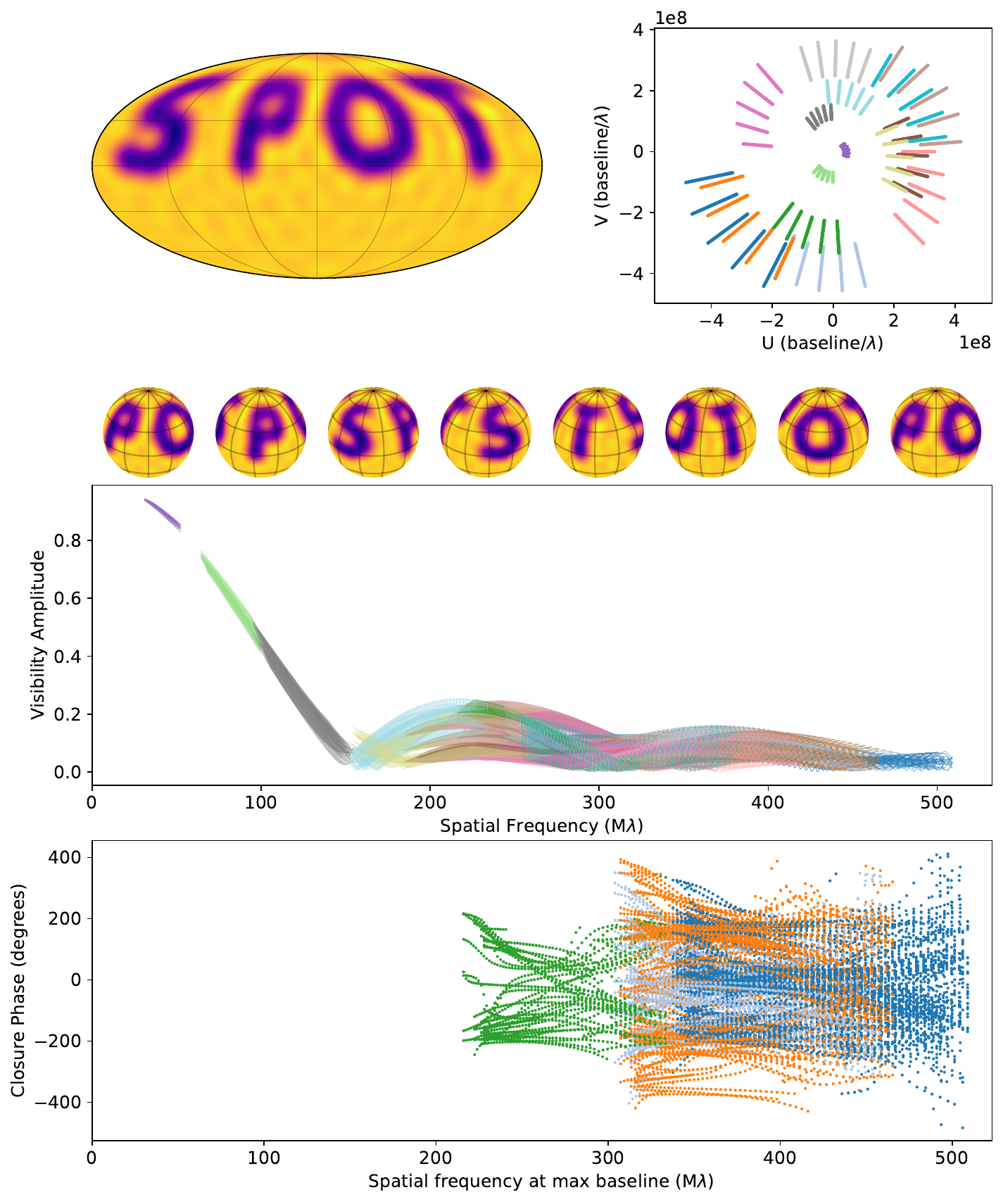} 
        \caption{The stellar rotation synthesis \texttt{SPOT} problem for a hypothetical observation with the \ac{chara} Array. We paint the word \texttt{SPOT} on the northern hemisphere of the star using an $l_{\mathrm{max}}=15$ in spherical harmonics, following \citep{luger2021b} (shown in a Mollweide projection at top left). The star is given a $60^{\circ}$ inclination, and 8 equally-spaced rotational phases are `observed' (2nd panel from top). Each observation is assumed to have the same $uv$ coverage, using all 6 telescopes to produce 15 independent baselines (top right, each baseline is assigned a color). The interferometric visibility amplitudes are colored corresponding to the baseline and each rotational phase is overplotted. The 10 independent closure phases are shown (bottom panel) and are colored corresponding to the maximum baseline in each triangle. \href{https://github.com/shashankdholakia/analytic-interferometry-paper/blob/main/src/scripts/spot_map_plot_chara.py}{\faGithub}}
        \label{fig:spotmap_chara}
    \end{centering}
\end{figure}

We use our \texttt{harmonix} model to compute two data vectors $\mathbf{d_{\mathrm{vis}}}$ and $\mathbf{d_{\mathrm{cp}}}$, corresponding to the visibility amplitudes and closure phases. The simulated spot map, along with the simulated $uv$ coverage, visibilities and closure phases are shown in Figure~\ref{fig:spotmap_chara}. 

\begin{figure}[ht!]
    \script{spot_map_plot_chara_optimization.py}
    \begin{centering}
        \includegraphics[width=\linewidth]{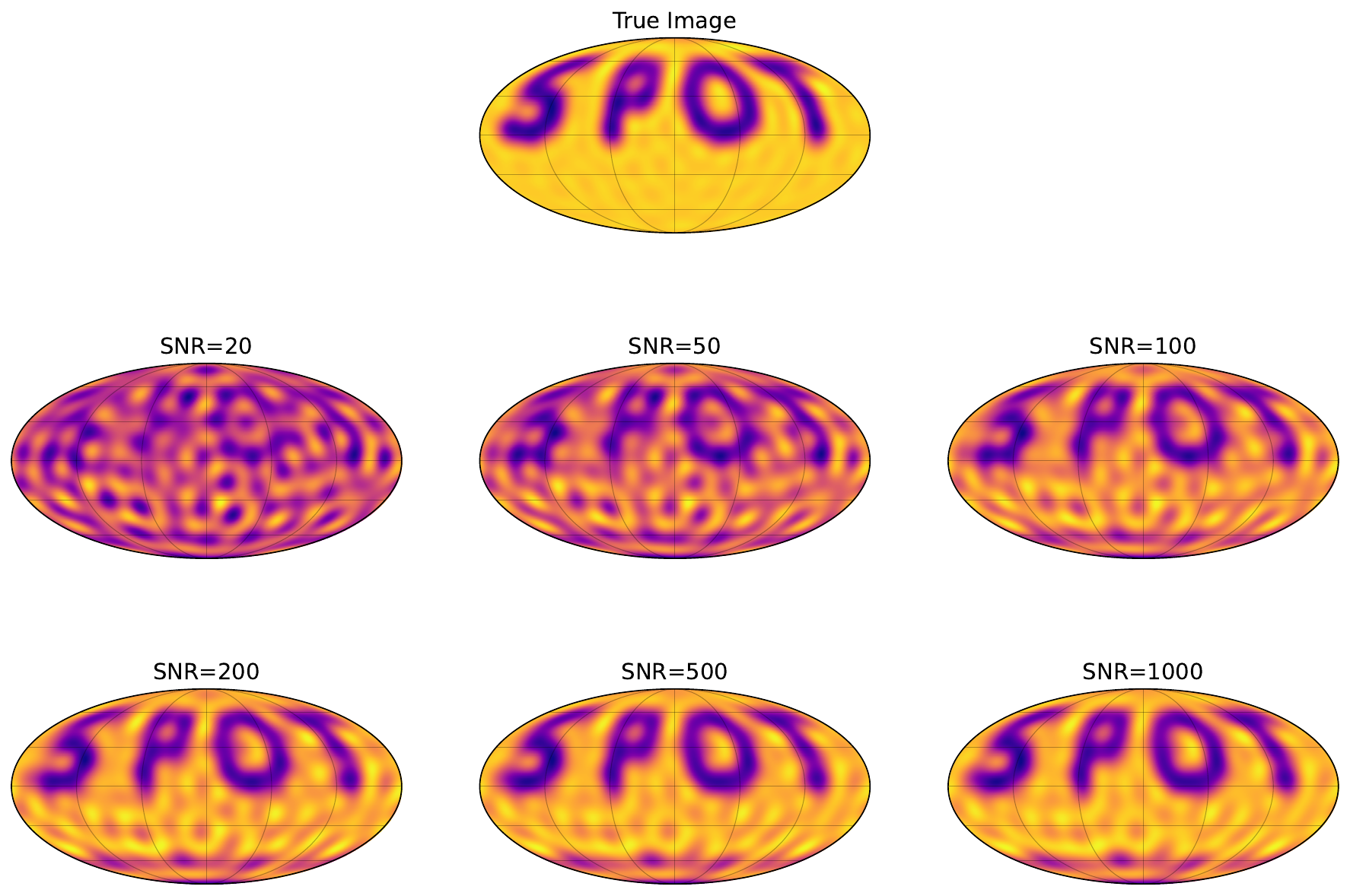} 
        \caption{Recovered surface maps at six different values of \ac{snr} for the \texttt{SPOT} problem with the CHARA Array's visibility amplitudes and closure phases without regularization. The star is assumed to have a known inclination of $60^{\circ}$, and as a result, latitudes below $-60^{\circ}$ are never visible. \href{https://github.com/shashankdholakia/analytic-interferometry-paper/blob/main/src/scripts/spot_map_plot_chara_optimization.py}{\faGithub}}
        \label{fig:chara_optimization}
    \end{centering}
\end{figure}

We then attempt to recover the \texttt{SPOT} map from the simulated data in the presence of a fixed amount of Gaussian noise using gradient-descent optimization. We start by injecting uncorrelated Gaussian noise with standard deviation $\sigma_{\mathbf{d}}$ into both the visibilities and closure phases. We define a loss function based on the \ac{mse} for both the visibility and closure phases. Because closure phases are an angular quantity, they wrap at multiples of $\pi$ and can cause problems when taking the \ac{mse} directly. We define an angular \ac{mse}, where we take the difference between the model and the data, wrap the resulting phase into the range $[-\pi, \pi)$ and square the result before taking the mean. Starting from a uniform surface map (with only the $l=0$ coefficient being nonzero), we use the \texttt{Adam} optimizer \citep{kingma2014} as implemented in \texttt{optax}, taking 4000 steps with a learning rate of $10^{-4}$. The results of this exercise for 6 different assumed Gaussian errors are shown in Figure~\ref{fig:chara_optimization}, where we define the \ac{snr} as $\sigma_{\mathbf{d}}=1/\mathrm{SNR}$.  

Next, we attempt to quantify the information and uncertainties relevant to the \texttt{SPOT} problem. First, we incorporate two-parameter limb darkening into the \texttt{SPOT} map with $\mathbf{u}=[0.1,0.1]$ to take into account flux lost towards the limb at each rotational phase. We define a log-likelihood function for visibilities and closure phases independently as shown in Eq.~\ref{eq:loglike}. We compute the second-order derivatives to construct the Fisher information matrix as shown in Eq.~\ref{eq:fim}, but instead with each of the spherical harmonic and limb darkening coefficients, totalling $257\times257$ terms. 

\begin{figure}[ht!]
    \script{spot_map_plot_chara_optimization.py}
    \begin{centering}
        \includegraphics[width=\linewidth]{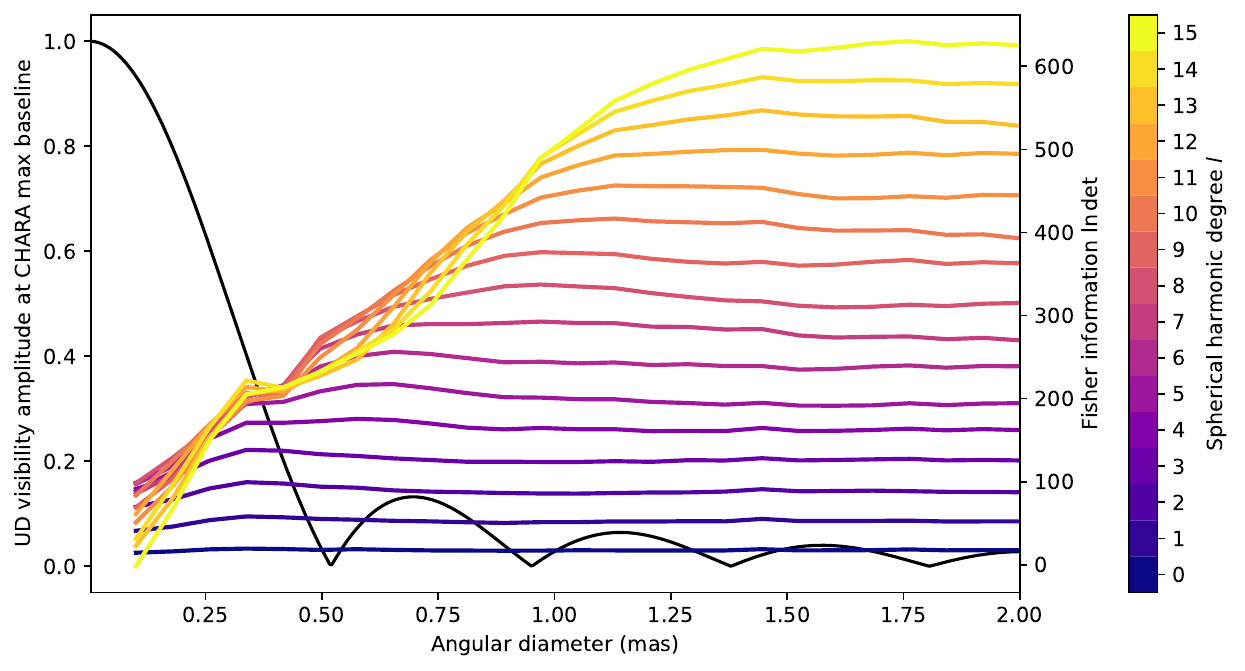} 
        \caption{Information content for each spherical harmonic degree $l$ as a function of stellar angular diameter for the CHARA \texttt{SPOT} problem. The black line indicates the visibility curve for a uniform disk star with a given angular diameter with uniform $u, v$ coverage out to the maximum baseline, and is meant as a reference for which null is reached for a given angular diameter. The information content is plotted as the natural log of the determinant of the block matrix containing only terms of a given degree $l$, color coded in ascending order. For each $l$, the information saturates at some maximum baseline, providing an effective bandlimit for a given angular diameter on an interferometric array. \href{https://github.com/shashankdholakia/analytic-interferometry-paper/blob/main/src/scripts/chara_sim_radius.py}{\faGithub}}
        \label{fig:chara_radius_information}
    \end{centering}
\end{figure}

Next, to compare the information from multiple experiments, we must establish an optimality criterion. One option is to use the variances on each term in the covariance matrix as shown earlier in Section~\ref{sec:angdiam}. However, even if this \acl{fim} is mathematically invertible to construct the covariance matrix, the resulting variances may be unrealistic, especially if certain coefficients contain little to no information. As a result, we use the D-optimality, which uses the determinant of the Fisher information matrix. In Figure~\ref{fig:chara_radius_information}, we show the results of an experiment varying the stellar angular diameter of the \texttt{SPOT} star to demonstrate the additional information as the star becomes better resolved. To do this, we subdivide the \ac{fim} into block matrices for each spherical harmonic degree $l$ and take the determinant of each submatrix. This gives a relative measure of the information added at each spatial scale on the star as we increase the angular diameter of the star. 

Lastly, we perform an experiment to demonstrate the concept of stellar rotation synthesis. Analogous to Earth rotation synthesis in interferometry, where information can be added to the 2 dimensional Fourier transform of a scene by using the Earth's rotation, adding interferometric data while allowing the \textit{star} to rotate, even by a very small amount, should add information on the 3 dimensional spherical harmonic modes representing the stellar intensity. It is already common practice in stellar interferometry to represent the star as a rotating 3D surface  \citep[e.g][]{roettenbacher2017, martinez2021}. However, the spherical harmonics being the eigenbasis for rotations on the sphere, it is natural to investigate the information theory of rotation synthesis in this basis. 

Continuing the \texttt{SPOT} model above, we generate synthetic visibilities and closure phases for a star with a rotation period of 5.1 days (the same as that of $\epsilon$ UMa) over 6 nights and with 5 different brackets over 4 hours at different position angles. We then investigate two scenarios: one in which data from each night is binned into a single, high SNR snapshot, and another in which we treat each bracket individually as part of an ensemble of visibilities and closure phases. Note that in both of these cases, we do build a full 3D model of the star using \texttt{harmonix}, but that in the latter case, we have a higher cadence over the course of the night. 

In each of these cases, we feed these simulated observations into \texttt{harmonix} and construct the \ac{fim} as above. We find that the $\ln{\det{\mathrm{FIM}}}$ is higher by $9\%$ and the covariance trace is higher for the binned observations by 7 times, implying a significant improvement in precision when each bracket is used individually. The structure of the \ac{fim} is also noticeably different. Similar to high-cadence photometry and Doppler imaging, we hypothesize that having even a few observations at higher cadence concentrates information on the higher-order modes by resolving small-scale shifts in starspot positions, and makes these modes more non-degenerate. 







\newpage
\subsection{Simultaneous photometry} \label{sec:simphotometry}

Due to the compatibility of \texttt{harmonix} with \texttt{jaxoplanet} \citep{jaxoplanet} (see Section~\ref{sec:harmonix} for further details), it is possible to simulate a space-based light curve obtained contemporaneously with interferometry, as done in a study by \citet{roettenbacher2017}. Due to the additive properties of Fisher information, we attempt to study this case to understand the information added by photometry, whether any degeneracies are broken, and how much overlap exists between observations.

\begin{figure}[ht!]
    \script{chara_sim.py}
    \centering
    \begin{tabular}{cc} 
        \includegraphics[width=0.45\textwidth]{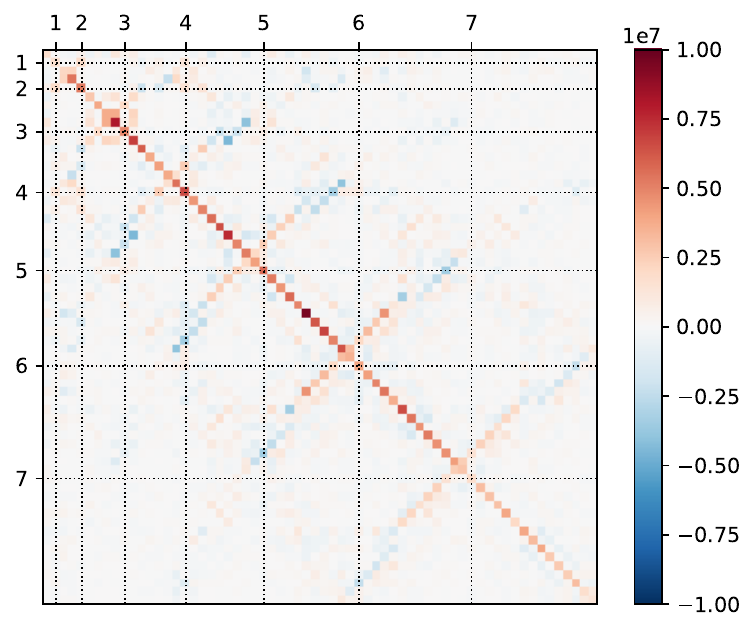} & 
        \includegraphics[width=0.45\textwidth]{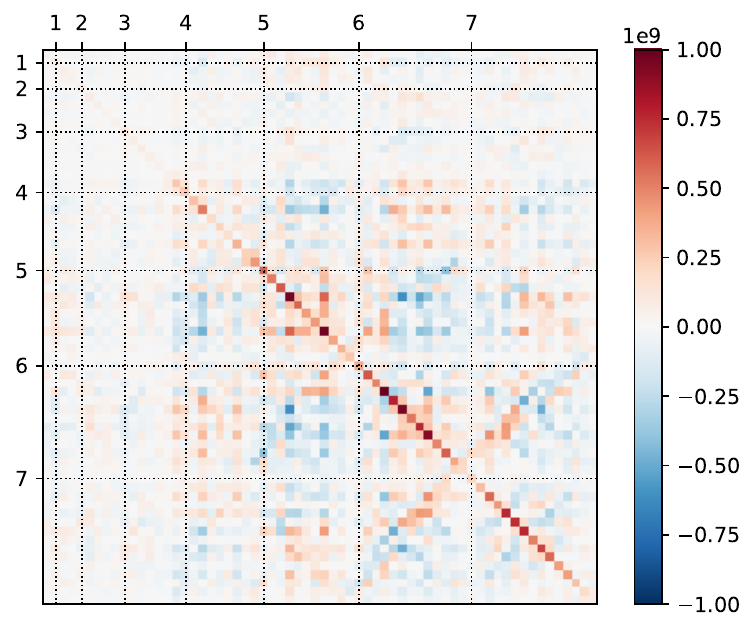} \\
        (a) visibility & (b) closure phase \\
        \includegraphics[width=0.45\textwidth]{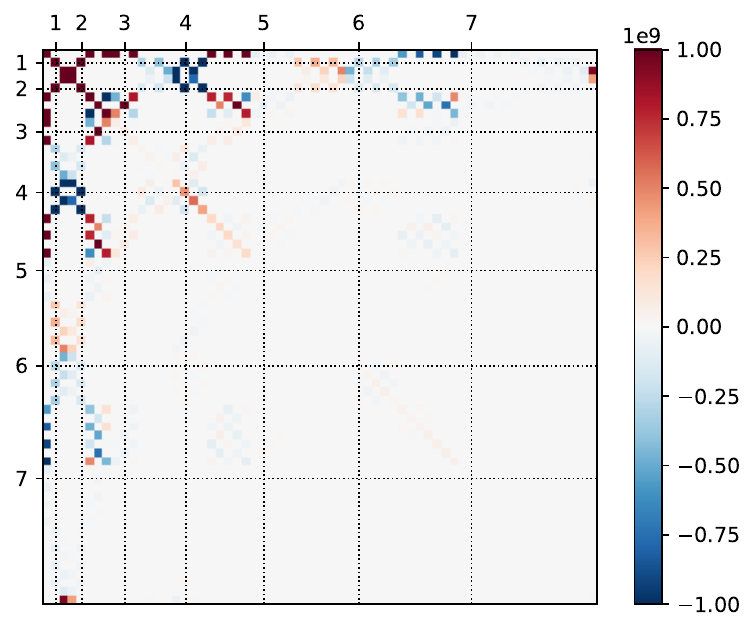} & 
        \includegraphics[width=0.45\textwidth]{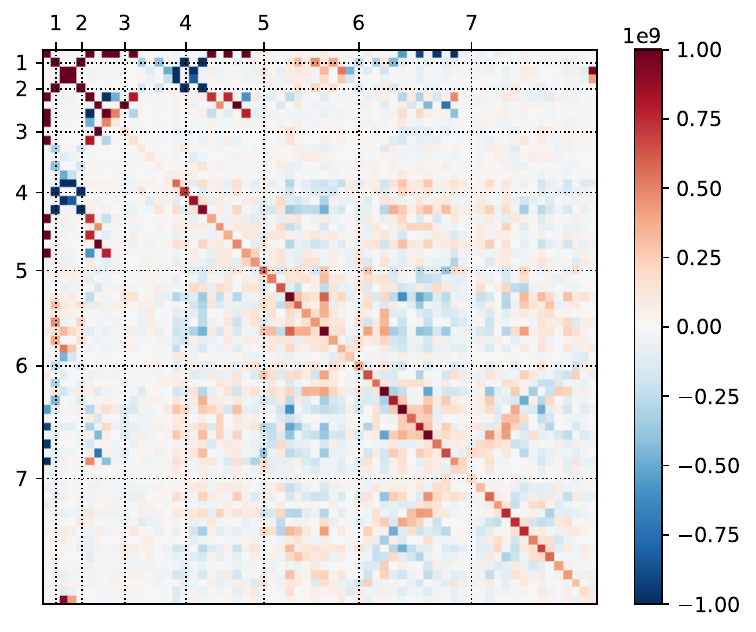} \\
        (c) light curve & (d) total
    \end{tabular}
    \caption{Fisher information matrices of the CHARA \texttt{SPOT} mapping problem with simultaneous photometry truncated to the first 7 spherical harmonic degrees for clarity. The origin (bottom left of each panel) is the $Y_{0,0}$ coefficient, and towards the top right show information on progressively higher degree coefficients. Note that the visibility amplitude (top left panel) is shown with a different linear normalization, since the information provided by visibility alone is much lower than the other data. At bottom right is the total \ac{fim}, when visibility amplitudes, closure phases, and simultaneous space-based photometry are all fit jointly. \href{https://github.com/shashankdholakia/analytic-interferometry-paper/blob/main/src/scripts/chara_sim_photometry.py}{\faGithub}}
    \label{fig:FIM}
\end{figure}

We start by simulating a rotational light curve with a cadence of 216 seconds and a duration of 5 days, similar to the 200 second cadence of Cycle 5+ of \ac{tess} \acp{ffi}. We construct the \ac{fim} of these data with respect to the same surface map coefficients as in Section~\ref{sec:rotsynthesis}, assuming a point-to-point scatter of $10^{-4}$. The results of this experiment, truncated to the first 7 spherical harmonic degrees for clarity, along with the \ac{fim} from the visibility amplitudes, closure phases, and all three techniques simultaneously, are shown in Figure~\ref{fig:FIM}.

The results of this experiment broadly agree with the finding of \citet{luger2021a} that the information content of light curves is largely confined to large scale structures and vanishes for higher-degree modes. Notably though, the high precision of space-based photometry contributes to \textit{very} high information on these low-degree modes; while the mode with the highest information using both visibilities and closure phases is roughly $10^8$, the light curve alone has modes with information of order $10^{11}$ (see Fig.~\ref{fig:FIM}). Furthermore, we find that interferometry with \ac{chara} \textit{lacks} information for these low-degree modes, making the information added by the two techniques highly complementary. The $\det{\mathrm{FIM}}$ (D-optimality as explained in Section~\ref{sec:rotsynthesis}) is two orders of magnitude higher when the photometry is added to the visibilities and closure phases. While inverting the \ac{fim} does not result in variances that are less than unity (as would be expected if the information on the surface map is complete enough for the inverse problem to be well-posed), the maximum error on the modes is nevertheless also reduced by 20 times. 

\subsubsection{Intensity interferometry} \label{sec:intensityinterferometry}

Intensity interferometry is an alternate technique for optical stellar interferometry first pioneered by \citet{hbt1956} and used to measure the angular diameters of 32 stars using the Narrabri Stellar Interferometer \citep{brown1974b}. It relies on the intensity signal in the quasimonochromatic regime, and measures correlations between these intensity fluctuations between pairs of telescopes. Intensity interferometry has limitations in signal-to-noise ratio and lacks phase information, so it is more difficult to use for stellar surface imaging. 

There has been a recent revival in intensity interferometry due to their complementary use case alongside gamma-ray observations with \acp{iact} \citep{hess2024, magic2024, veritas2022, acharya2024}. Intensity interferometry with \acp{iact} has two main advantages compared to current amplitude interferometers for stellar surface imaging. Firstly, the visibility amplitudes are self-calibrated without needing to observe calibrator stars, allowing a higher cadence of observations. Secondly, there is no requirement to physically combine the light, so it is in theory easier to construct arrays with much denser and/or kilometric baselines to allow imaging of more distant stars. 

\begin{figure}[ht!]
    \script{spot_map_plot_iact.py}
    \begin{centering}
    \includegraphics[width=\linewidth]{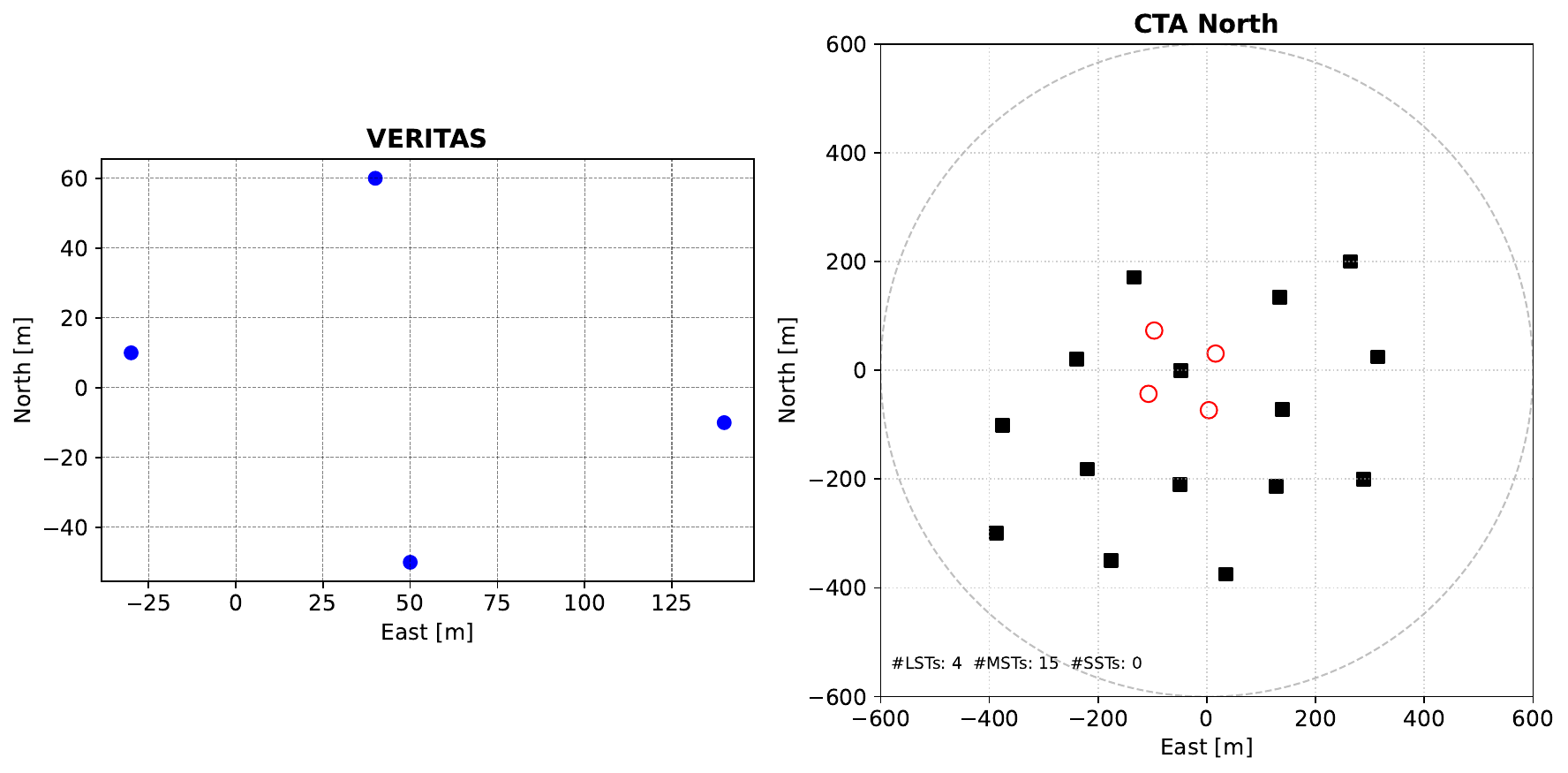}
        \caption{Reflector configurations for our two simulated intensity interferometers based on \ac{veritas} and \ac{cta} North with the proposed `omega' configuration. \href{https://github.com/shashankdholakia/analytic-interferometry-paper/blob/main/src/scripts/spot_map_plot_iact.py}{\faGithub}}
        \label{fig:iact_arrays}
    \end{centering}
\end{figure}

We simulate the \texttt{SPOT} problem for intensity interferometry in order to address the possibility of stellar surface imaging with this technique. We use two examples of intensity interferometers: the \ac{veritas} array, which contains 4 reflectors with a maximum baseline of 173\,m and the proposed \ac{cta} North in its omega configuration \citep{maier2019performance}, with 15 medium scale reflectors and 4 large scale reflectors with a maximum baseline of roughly 800\,m. In these simulations, like above, we assume that each individual reflector can be modelled as a point aperture and do not consider effects from the large sizes of each reflecting dish. We show the reflector positions for both \ac{veritas} and \ac{cta} North in Figure~\ref{fig:iact_arrays}.

\begin{figure}[ht!]
    \script{spot_map_plot_veritas.py}
    \begin{centering}
        \includegraphics[width=\linewidth]{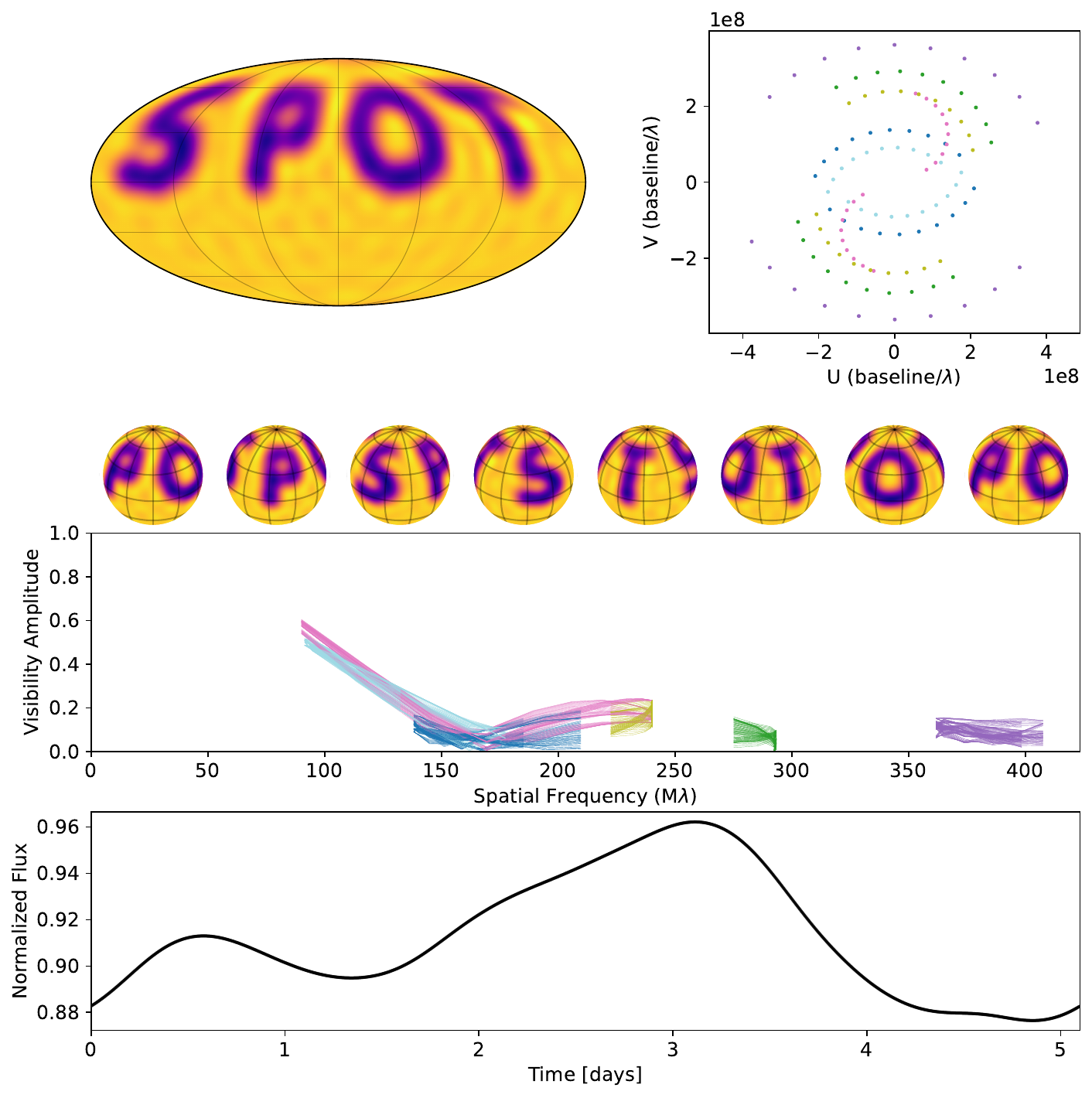}
        \caption{The stellar rotation synthesis \texttt{SPOT} problem for a hypothetical observation with the \ac{veritas} Array. The star is identical to the simulation above with \ac{chara}, but the telescope setup is different--there are 4 telescopes, 6 baseline combinations, only one wavelength, and due to the nature of intensity interferometry, no phase data. \href{https://github.com/shashankdholakia/analytic-interferometry-paper/blob/main/src/scripts/spot_map_plot_veritas.py}{\faGithub}}
        \label{fig:spotmap_veritas}
    \end{centering}
\end{figure}

We start by using a similar procedure as before, obtaining the $u,v$ positions of the fiducial star $\epsilon$ UMa over 10 hour angles per night at the latitude of the \ac{veritas} in Arizona for a central wavelength of $650$nm. The 4 \ac{veritas} reflectors yield 6 baseline combinations. We repeat this procedure for 8 `nights' which are equally spaced in rotational phase over the \texttt{SPOT} star's rotation period (see Figure~\ref{fig:spotmap_veritas}). 

In addition to the intensity interferometer, we also assume we have simultaneous space-based photometry, as described in Section~\ref{sec:simphotometry}. We construct the data vector for the visibilities as described in Section~\ref{sec:rotsynthesis}, injecting three different levels of Gaussian random noise into the visibility curve and of $\sigma_{\mathrm{lc}} = 10^{-4}$ into the light curve. 

We then perform gradient-descent minimization to attempt to recover the best-fitting surface map from this simulation. We define a loss function as above using the sum of the \ac{mse} for the visibility and light curve. Because intensity interferometry contains analytic degeneracies due to the lack of phase information, we use total-variation (TV) regularization with $\lambda=10^{-7}$, which favors piecewise smooth surfaces and penalizes excessive high frequency variations.

\begin{figure}[ht!]
    \script{spot_map_plot_iact_optimization.py}
    \begin{centering}
    \includegraphics[width=\linewidth]{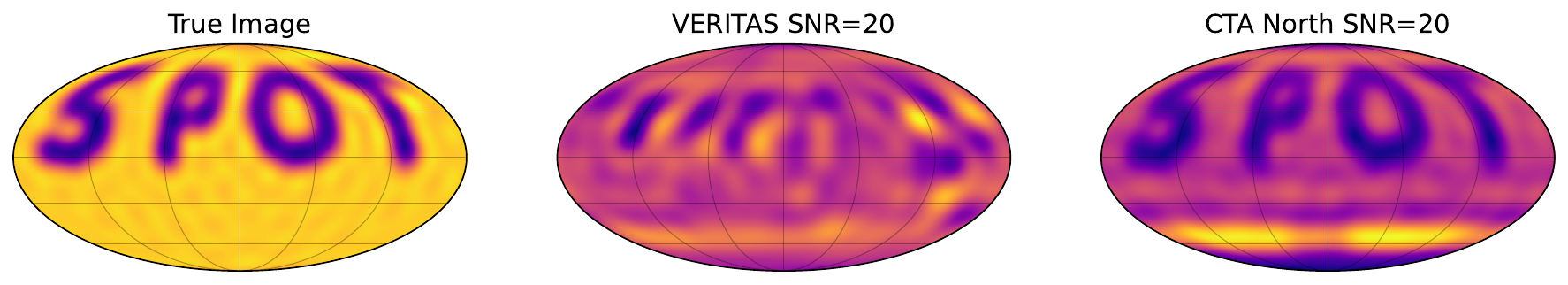}
        \caption{Recovered surface map for the \texttt{SPOT} problem with \ac{veritas} and \ac{cta} North intensity interferometers using total variation (TV) regularization. \href{https://github.com/shashankdholakia/analytic-interferometry-paper/blob/main/src/scripts/spot_map_plot_iact_optimization.py}{\faGithub}}
        \label{fig:iact_optimization}
    \end{centering}
\end{figure}

The results of this optimization are shown in Figure~\ref{fig:iact_optimization}. We find that for \ac{veritas}, despite adding regularization and even at relatively high SNR, we are unable to reconstruct a legible \texttt{SPOT} map, although certain features above the equator are recovered. We can attribute this to the small number of available baselines, even lower than \ac{chara} and without closure phases. 

To investigate whether future \acp{iact} with more dense sampling of the $u,v$ plane could overcome the inherent challenges of intensity interferometry, we also repeated the simulation and gradient-descent procedure with the proposed \ac{cta} North in its proposed omega configuration, along with simulated space-based photometry as before. We assume the same central wavelength and hour angles per night, but a total of 19 reflectors yields 171 baseline combinations, far more than existing long-baseline optical stellar interferometers. The recovered \texttt{SPOT} map is shown in Figure~\ref{fig:iact_optimization} on the right. We find that the \texttt{SPOT} map is recovered easily albeit with some banding artifacts towards the lower pole.



\section{Implementation in harmonix}
\label{sec:harmonix}

We implement the closed-form solutions in Section~\ref{sec:solution} in the open-source Python package \texttt{harmonix}, which is written using the \textsc{Jax} framework for \ac{jit} compilation and automatic differentiation. The implementation is designed on top of the \texttt{jaxoplanet.starry} package, which allows photometric modelling of bodies with spherical harmonics, including rotation and occultations \citep{jaxoplanet}. The goal of \texttt{harmonix} is to maintain cross-compatibility with \texttt{jaxoplanet} and other similar packages to build towards a suite of software for modeling and fitting photometry, Doppler imaging, and interferometry together. 

Users can construct a \texttt{Harmonix} object by passing in a \texttt{jaxoplanet.starry.Surface} object, (which contains relevant parameters such as the inclination, obliquity, rotational period, and surface map coefficients) as well as an angular radius in \acl{mas}. The resulting object's \texttt{model} function takes a set of $u, v$ coordinates (dimensionless; projected baseline separation over wavelength) and a time, and returns the complex visibility for each $u, v$ coordinate for the visible surface of the star at the time. With automatic vectorization in \textsc{Jax}, this function can be extended to simultaneously compute the visibility at any number of times, baseline separations, or wavelengths. Also included are several utility functions for converting the complex visibility function into observed quantities such as visibility amplitudes and closure phases.

\texttt{harmonix} is also designed to be user-friendly with an object-oriented \ac{api} built on the packages \texttt{equinox} \citep{kidger2021} and \texttt{zodiax} \citep{Desdoigts2024} for \ac{jit} compiling and differentiating through objects in \textsc{Jax}. This also makes it easy to use with several libraries for inference and optimization in \textsc{Jax} such as \texttt{optax} \citep{optax} and \texttt{numpyro} \citep{phan2019}.

Lastly, we have included several unit tests benchmarking \texttt{harmonix} against existing methods where possible to ensure that they agree. First, we test \texttt{harmonix} against the simplest subcase, that of a uniform disk, which has an analytic visibility of an Airy function, and ensure that the results match to within floating-point precision. Next, we benchmark the \texttt{harmonix} model against an equivalent implementation using \texttt{scipy} \citep{scipy} to ensure numerical stability in \textsc{Jax}. Incidentally, we find the \textsc{Jax} versions of Bessel and spherical Bessel functions, now deprecated, are numerically unstable, and reimplement these functions ourselves. We then test the math by comparing the integrals presented in Section~\ref{sec:solution} against the same integrals computed with symbolic integration in \texttt{Mathematica} up till $l_{\mathrm{max}}=10$, and find that they agree within $10^{-5}$ everywhere, although the \texttt{Mathematica} solutions without simplification have numerical instabilities which prevent agreement to floating-point precision.

Lastly, as an end-to-end test of the \texttt{harmonix}, we render a body with a surface map projected into 2D into a $400\times400$ pixel grid at various viewing orientations and compute the visibility function using the \ac{dftm}. We compare this to \texttt{harmonix} by using a spherical harmonic transform to render the surface at a resolution of $l_{max}=15$, and ensure that they agree within the expected precision of the discretization.





\section{Discussions}
\label{sec:discussions}

In Section~\ref{sec:simphotometry}, we show that contemporaneous space-based photometry can add information about stellar surfaces which is complementary to long baseline optical stellar interferometry. While such photometry is not capable of adding information on modes of higher degree, it is able to provide a large amount of information on low degree modes due to its high precision, whereas interferometry provides comparatively little information on these modes. The resulting order of magnitude improvement on the precision of the surface map parameters may help bringing nearby main sequence stars into view with modern interferometers such as \ac{chara}. 

The experiment shown in Section~\ref{sec:intensityinterferometry} shows that future intensity interferometers such as \ac{cta} with many reflectors and hence dense sampling of the $u, v$ plane is in fact sufficient to overcome some of the inherent limitations in intensity interferometry, such as the lack of phase information and the relatively low \ac{snr}, especially in conjunction with simultaneous photometry. This experiment assumed a relatively high \ac{snr} of 20 to demonstrate good results for the surface map, which would be difficult to achieve in a single observation for many main sequence stars; limited spot lifetimes would prevent continuous long term coverage of a single surface map. However, we envision that the advantages of future \ac{iact}-based intensity interferometers may lie in detailed spot maps of chemically peculiar stars. Their early type better suits the wavelength dependence of intensity interferometers and their spots are in many cases extremely stable over decades. This will be especially the case if the central wavelength of these intensity interferometers can be chosen to coincide with specific abundance features. 

\subsection{Limitations}
There are a few important limitations to the work presented in this paper, both in the implementation of the algorithm and in the information theory. Firstly, we rely heavily on the spherical harmonics as the basis for the solutions, which can have limitations at very high angular resolutions, both in terms of memory and numerical precision. Unlike \citet{starry2019}, which notes precision loss for $l_{\mathrm{max}}>30$, our solution does not rely on ratios of factorials. Nevertheless, evaluating Bessel functions at very high orders is numerically challenging and may be the limiting factor in the precision at high spherical harmonic degrees. In addition, the memory required to compute the rotation matrices as implemented in \texttt{starry} \citet{starry2019} can balloon at high degrees. We note that there exist highly optimized \textsc{Jax} implementations of spherical harmonic rotations out to $l_{\mathrm{max}}>1000$, which we can use to greatly reduce these limitations. A similar highly-optimized Bessel function implementation would be very useful in many fields of physics, but does not yet seem to exist in the \textsc{Jax} ecosystem. 

The spherical harmonic basis, while a natural basis to represent rotation and the information theory in this paper, also has more fundamental limitations that may not be overcome by improvements in implementation. Firstly, the spherical harmonics are a smooth basis and hence not well-suited to modeling sharp discontinuities in stellar surfaces, such as hard-edged spots or faculae. Other similar bases, such as spherical wavelets \citep{price:s2wav} or combinations of delta functions, could be used to better represent fine-grained surface details, but would require closed-form solutions for visibility as presented in this paper. Furthermore, as noted by \citet{starry2019}, the spherical harmonic basis has no natural way of enforcing that maps are nonnegative everywhere, as would be expected for a stellar surface. In some cases, it may be practical to perform inference in a pixel basis with non-negative priors on each pixel and use packages such as \texttt{s2fft} to transform into the spherical harmonic basis to quickly compute interferometric observables. 


There are also other physical extensions to \texttt{harmonix} that may need to be implemented to fit real data of nearby main-sequence stars, such as multiwavelength surface maps and spectrally-variable surface features, surface evolution of spots, differential rotation, and non-sphericity. Some of these already have implementations in \texttt{jaxoplanet} and only need minor modifications to use them in \texttt{harmonix}, such as multiwavelength maps. Non-sphericity of the star due to tides or rapid rotation can be taken into account using an affine transformation of the stellar surface without modification of the above integrals as shown in \citet{dholakia2022}, as long as the star is well-represented as a spheroid or triaxial ellipsoid.  Others, like differential rotation and time evolution of spots, are somewhat harder to implement. Again, a pixel-basis inference where latitudes are allowed to have different rotational velocities, as well as allowing time evolution of surface spherical harmonic modes, for instance with the use of Gaussian processes, may be necessary for these problems. 

We also rely heavily on Fisher information to parametrize the achievable precision of observations. While Fisher information gives useful insights into the relative precision of simulated observations, we note certain limitations in extending these results to real observations. Firstly, we use the true, as opposed to the \textit{observed} Fisher information, by directly computing the Hessian at the known parameters of the \texttt{SPOT} map. In reality the maximum likelihood solution will be different from the true solution due to noise and biases. Fisher information also assumes the posterior is well-represented as a Gaussian, which is accurate only in the limit of infinite data. As a result, we caution against interpreting covariances estimated using this method directly as uncertainties in real observations. 

We also have not marginalized over certain parameters in this analysis, such as the stellar inclination, obliquity, or rotation period. Real data necessitates a fully Bayesian approach, with isotropic priors on the inclination and obliquity, constraints on the rotation period from existing observations, and perhaps regularization on the surface map coefficients implemented explicitly through a prior. Then, posterior distributions on the surface map can be estimated with the use of \ac{hmc} or similar gradient-based samplers.

\section{Conclusions} \label{sec:conclusions}

We have presented a new analytic solution for the visibility function of a rotating star with its surface intensity expressed using spherical harmonics. Using this model, we have investigated the information theory of stellar surface mapping with interferometry. We showed that a long-baseline stellar interferometer such as \ac{chara} is capable of accurately mapping the surfaces of rotating stars without assumptions in the form of regularization or priors. We introduced the concept of stellar rotation synthesis in which information on spherical harmonic modes representing the star's surface map is added by observing closely spaced rotational phases. We find that adding photometric information in the form of a space-based rotational light curve complements the information added by optical long-baseline interferometry. Lastly, we compare the existing intensity interferometer \ac{veritas} to the proposed \ac{iact} to show their performance for stellar surface mapping and demonstrate that future \ac{iact}-based intensity interferometers may be able to map some stars accurately. 

A significant goal of this work is to add \texttt{harmonix} to a suite of mutually compatible software in \textsc{Jax} to map the surfaces of stars with different observations. Software in this ecosystem such as jaxoplanet \citet{jaxoplanet} (uses the spherical harmonic basis) and \texttt{spotter} \citet{spotter} (uses a pixel basis) already exist for modeling light curves in advanced ways. A future work will address the translation of \texttt{starry}'s existing Doppler imaging framework based on spherical harmonics into a \textsc{Jax} package; \texttt{spotter} already contains utilities to do the same using pixels. A package in \textsc{Jax} similar to \texttt{SURFING} \citep{roettenbacher2016} that uses pixels on a spheroidal star would also be useful. Another technique that could benefit from inclusion in this ecosystem is Zeeman Doppler imaging.  

Having automatically differentiable software that are compatible at some level is not merely a convenience with fitting and inference routines. With hundreds or thousands of parameters needed to model the surfaces of stars at high resolutions, standard inference techniques such as affine-invariant ensemble sampling can silently fail to converge \citep{huijser2015}, and others such as nested sampling would take impracticably long. In our case, the use of gradient-based inference techniques such as \ac{hmc} is needed to create posterior distributions on surface maps. This requires that the entire model, starting with the parameters representing the surface intensity, and ending with the various observables, be end-to-end differentiable, which is made practical with automatic differentiation. Notwithstanding, there are also other reasons to prefer such an ecosystem in \textsc{Jax} or related machine-learning packages, including the ability to perform gradient-based optimization, \ac{jit} compilation, and a well-developed set of inference routines. 

Till date in the literature, directly imaging spots on rotating stars has been restricted to red giants \citep{roettenbacher2016, roettenbacher2017, martinez2021, anugu2024}, with one marginal result in a main sequence star \citep{roettenbacher2022}. Of course, we would like to be able to identify active latitudes, explore differential rotation, and understand starspot properties in order to do \textit{physics}; learning about the magnetic dynamos and interiors of stars, in addition to solving problems in other areas of astrophysics such as exoplanet detection and characterization. We have shown in this paper that submilliarcsecond imaging is within the reach of existing interferometric facilities such as \ac{chara}, especially when performed jointly with photometry. Synergies between interferometry and Doppler imaging are likely to exist as well and will be explored in a future work. We propose that with a combination of joint inference with photometry, spectroscopy, and interferometry, as well as the use of more advanced gradient-based Bayesian optimization and inference, that it will be possible to create detailed maps of nearby main-sequence stars for the first time and begin to use interferometry to address the above goals. 

Even with these improvements, interferometry with current facilities may not be able to study the surface maps of main-sequence stars at a statistical level. Proposed ground-based interferometers such as the Big Fringe Telescope \citep{bft2024}, or future space or lunar-based interferometers may be able to achieve interferometric images of the surfaces of main-sequence stars at a large scale and potentially also image their (unresolved) exoplanets directly \citep{life2024}. 

The information theory presented in this work may also be useful to the planning of these proposed interferometers. It is common in many fields \citep[e.g cosmology;][]{amara2011, virey2007} to have a quantitative `figure of merit' when planning large missions or surveys which is based on Fisher forecasting certain parameters of interest. Recently in the broader field of optical interferometry, the \acs{toliman} mission took this a step further by using the Fisher information as an optimality criterion in the construction of their diffractive pupil, minimizing the covariance over the parameter of interest marginalized over all other parameters \citep{Desdoigts2024}. In the example of long-baseline optical stellar interferometry, if we define the determinant of the Fisher information matrix or trace of the covariance matrix on the simulated surfaces of nearby stars as an optimality criterion, it will be possible to optimize the positions of each telescope, bandpasses, spectral resolution or other aspects of the interferometer to maximize its scientific return. 





\newpage
\section*{Acknowledgments}

We wish to thank Keaton Burns and Geoff Vasil for their assistance and advice in the derivations.

SD and BJSP were funded by the Australian Government through the Australian Research Council DECRA fellowship DE210101639.

We acknowledge and pay respect to the traditional owners of the land on which the University of Queensland and Macquarie University are situated, upon whose unceded, sovereign, ancestral lands we work. We pay respects to their Ancestors and descendants, who continue cultural and spiritual connections to Country.

\software{
Astropy \citep{astropy:2022}, \textsc{NumPy} \citep{numpy}; Matplotlib \citep{matplotlib}; \textsc{Jax} \cite{jax}; \texttt{jaxoplanet} \citep{jaxoplanet}; \texttt{numpyro} \citep{phan2019}; \texttt{optax} \citep{optax}; \texttt{scipy} \citep{scipy}
}

\bibliography{bib}
\end{document}